\documentclass[a4paper, 11pt]{article}

\usepackage{graphicx}
\usepackage{subcaption}
\usepackage{geometry}
\usepackage[version=4]{mhchem}
\usepackage[round]{natbib}
\usepackage{hyperref}
\usepackage{parskip}
\usepackage{amsfonts}
\usepackage{booktabs}
\usepackage{authblk}

\title{Archetypal Microbiome Profiles as Indicators of Nitrous Oxide Emission States in Activated Sludge}
\author[1,2]{Cheng Chen\thanks{Corresponding author: cheng.chen@eawag.ch}}
\author[3]{Marcello Seppi}
\author[3]{Samir Suweis}
\author[1]{Andreas Froemelt}
\author[1,2]{Eberhard Morgenroth}
\author[1]{Andreas Scheidegger}
\author[1]{Carlo Albert}

\affil[1]{Eawag, Swiss Federal Institute of Aquatic Science and Technology}
\affil[2]{Institute of Environmental Engineering, ETH Zurich}
\affil[3]{Department of Physics and Astronomy ``G. Galilei'', University of Padua}
\date{}

\begin{document}

\maketitle

\begin{abstract}
    Nitrous oxide (\ce{N2O}) emissions from water resource recovery facilities (WRRFs) fluctuate over time and can arise from multiple microbial pathways, making source attribution and full-scale prediction difficult. The difficulty is compounded by the high dimensionality of activated sludge microbiomes, whose complex and dynamic community structure can obscure relationships with \ce{N2O} emission patterns. This study evaluated whether interpretable, low-dimensional representations of activated sludge microbiomes can be correlated with \ce{N2O} emission states. Temporal 16S rRNA gene amplicon profiles and \ce{N2O} emission metrics were collected from two full-scale WRRFs in Switzerland. Genus-level relative-abundance profiles were summarized using archetypal analysis (AA), which represents each sample as a convex combination of a small number of interpretable community profiles. In both WRRFs, three archetypes captured most explainable variation in community composition (63\%--73\%) and defined a simplex state space in which samples clustered near vertices and edges, indicating that community compositions were organized around distinct archetypal states and their mixtures. Without using emission labels while training, the archetypal state space aligned strongly with binary \ce{N2O} emission states: high-emission observations in both plants concentrated around a specific archetype, and temporal trajectories showed consistent high weights of this archetype during high-emission periods. Functional summaries suggested site-specific but pathway-relevant interpretations of the high-\ce{N2O} archetype. Temperature further structured the archetypal state space, indicating seasonal forcing of microbiome configurations associated with elevated \ce{N2O}. Overall, AA provides an interpretable framework to track microbiome regime shifts and may support operational tracking of high-\ce{N2O} emission states in full-scale WRRFs.
\end{abstract}


\section{Introduction}
Nitrous oxide (\ce{N2O}) is a potent greenhouse gas (GHG) that contributes substantially to climate change and ozone depletion \citep{ravishankara2009}. Water resource recovery facilities (WRRFs) are recognized as significant emission hotspots, as the biological wastewater treatment processes therein might lead to \ce{N2O} formation \citep{law2012,vasilaki2019}. In WRRFs, \ce{N2O} emissions can account for a large portion of the overall GHG footprint \citep{daelman2013,song2024}. This further underscores the urgency of understanding the emission mechanisms and developing mitigation strategies. 

It has been realized that the \ce{N2O} emission in the WRRFs displays seasonal and diurnal dynamics \citep{daelman2015,gruber2020,valk2022,roothans2025,froemelt2025}, and the \ce{N2O} in activated sludge processes arises primarily from three microbial pathways performed by two groups of bacteria: (i) nitrifier denitrification by ammonia-oxidizing bacteria (AOB); (ii) hydroxylamine (\ce{NH2OH}) oxidation by AOB; and (iii) heterotrophic denitrification, where \ce{N2O} is an obligate intermediate \citep{wunderlin2012,duan2020,xie2023}. Recent experimental work has further shown that the relative importance of these pathways can be disentangled under different operating conditions \citep{strubbe2026}. Meanwhile, the capacity of microbial communities for the consumption of \ce{N2O} depends on organisms carrying the nitrous oxide reductase gene \textit{nosZ}, which is now known to exist in two phylogenetic clades with distinct ecological distributions and substrate affinities \citep{hallin2018,qi2022,schacksen2024,laureni2025}. However, due to the complexity of emission patterns driven by different functional bacteria and the very high dimension of this functional space, it is still challenging to trace the active pathways at full-scale WRRFs, which further complicates the predictive quantification of \ce{N2O} emissions \citep{bellandi2020,vasilaki2020,gruber2022,ye2022,seshan2025}. Nevertheless, a comprehensive characterization of the microbial community structure in WRRFs is still expected to provide predictive insight into \ce{N2O} production and consumption dynamics.

Recent studies have therefore started to explicitly pair \ce{N2O} monitoring with microbial community characterization to identify microbial signatures of different emission regimes in activated sludge. \citet{castellano-hinojosa2018} linked temporal \ce{N2O} fluxes in four municipal WRRFs to the population dynamics of nitrifying and denitrifying organisms, showing that AOB abundance was positively correlated with biological \ce{N2O} generation across plants. Across three full-scale WRRFs, \citet{vieira2019} combined 16S rRNA gene amplicon sequencing with functional gene quantification and principal component analysis (PCA), demonstrating that \ce{N2O} emission factors covaried with genus-level community structure and the expression of denitrification genes (including \textit{nirK} and \textit{nosZ}). More recently, \citet{yan2024} followed an anoxic-oxic WRRF operated at low dissolved oxygen and low influent C/N over nine months and showed that \ce{N2O} emission factors were strongly associated with the relative abundance of \textit{Nitrosomonas} and \textit{Terrimonas}, as revealed by 16S rRNA sequencing, qPCR of nitrification/denitrification genes and principal coordinates analysis (PCoA). In contrast, lab-scale anoxic-oxic systems have sometimes reported weak direct links between community composition and emissions: \citet{guo2021} observed pronounced pH-dependent \ce{N2O} production but found no clear correlation between \ce{N2O} fluxes and potential nitrifiers/denitrifiers, suggesting that process conditions can modulate enzyme activities and intermediates more strongly than community composition \textit{per se}.

In systems with pronounced seasonal dynamics, ordination-based approaches have been used to connect emission states to shifts in community structure. In sequencing batch reactors (SBR), \citet{gruber2021} combined long-term online \ce{N2O} monitoring with 16S rRNA amplicons and non-metric multidimensional scaling (NMDS), and discovered that peak emission phases coincided with marked declines in nitrite-oxidizing bacteria (NOB) and filamentous taxa, consistent with a transition to an \ce{NO2-}-accumulating, high-\ce{N2O} state. \citet{valk2022} applied NMDS at the species level and identified 26 taxa positively and 28 taxa negatively associated with \ce{N2O} concentrations, implicating that nitrifiers and denitrifiers jointly shape emission regimes. \citet{kinnunen2025} analyzed parallel lines during the high emission phases using 16S amplicon and PCoA, and showed a clear community separation between the high and low \ce{NO2-}/\ce{N2O} lines, as well as a significant decrease in NOB, accompanied by changes in filamentous and denitrifying taxa. Together, these studies consistently associate \ce{NO2-} accumulation, NOB loss, and altered denitrifying taxa with transitions into high-\ce{N2O} emission states.

Beyond 16S-based surveys, meta-omics has been deployed to resolve the functional basis of these community shifts. \citet{roothans2025} coupled nearly two years of metagenome-resolved metaproteomics with \textit{ex situ} kinetics and full-scale operational data from a full-scale plant, and showed that seasonal \ce{NO2-}/\ce{N2O} peaks were driven by coordinated changes in AOB and NOB abundance and activity, with nitrifier denitrification emerging as the dominant \ce{N2O} production pathway under low-\ce{O2} control. Parallel works on sludge samples and enrichments derived from sludge have further emphasized the importance of \textit{nosZ}-carrying \ce{N2O} reducers: group-specific qPCR and meta-omics studies have revealed clade II–dominated \textit{nosZ} pools and identified specific \textit{Flavobacterium}- and \textit{Dechloromonas}-like populations as key \ce{N2O}-sinks \citep{kim2020,kim2022,qi2022}.

Collectively, these previous studies demonstrate that microbial community composition can be a critical determinant of \ce{N2O} emissions from WRRFs, and taxonomically classified community data can reveal microbial configurations associated with distinct \ce{N2O} emission regimes. However, they also show that the relationships are complex and are typically explored in a high-dimensional ordination space with a lack of explainability. This motivates the development of interpretable and low-dimensional representations of microbiomes that can be correlated with \ce{N2O} emission states.

Most prior studies rely on dimension-reduction techniques such as PCA, PCoA or NMDS to project communities into a low-dimensional ordination space. While powerful for visualization, these approaches yield abstract embedding axes that are linear combinations of taxa with positive and negative loadings; therefore, they cannot be directly interpreted as feasible community compositions. Archetypal analysis (AA), in contrast, summarizes each sample as a convex mixture (non-negative coefficients summing to one) of a small set of archetypes and reconstructs community profiles in a way that respects their relative-abundance structure \citep{cutler1994}.

This formulation has two advantages for our problem. Firstly, it enhances interpretability: archetypes correspond to extreme microbial profiles that can be examined in terms of specific taxonomic and functional markers, rather than abstract ordination axes \citep{ragozini2017,hes2023,meawad2025}; Secondly, the convex representation provides a natural way to track system dynamics, as temporal changes in activated sludge can be described as trajectories in the simplex of archetype \citep{keller2021}. This defines a low-dimensional, mechanistically interpretable state space within which emission regimes and operational interventions can be compared, making AA particularly suitable when the goal is to link microbial community structure to \ce{N2O} emission states and possible pathways, rather than solely to improve ordination fit.

In this study, we investigate whether distinct patterns in sludge microbial composition can serve as explainable indicators of \ce{N2O} emission levels in two full-scale WRRFs. Whereas short-term experimental studies, such as \citet{strubbe2026}, are designed to disentangle specific \ce{N2O} production pathways under controlled conditions, our focus is on long-term \ce{N2O} emission behavior at full scale. Specifically, we ask whether persistent microbiome configurations can be summarized by archetypes that reveal stable community regimes associated with \ce{N2O} emission levels. In this way, we use archetypes as an interpretable bridge between microbial community structure and possible emission pathways by examining how functional microbiome abundance profiles relate to \ce{N2O} emissions, while also tracking transitions in community profiles over time.

\section{Materials and methods}
\subsection{Sampling sites}
The study was conducted at two full-scale municipal water resource recovery facilities (WRRFs) in the canton of Zurich, Switzerland: the Zürich-Werdh\"olzli WRRF (hereafter ``Werdh\"olzli'') and the Jungholz WRRF in Uster (hereafter ``Uster''). Both facilities mainly treat domestic wastewater with minor industrial contributions and are designed for biological nitrogen removal. However, they differ notably in scale and process configuration.

Werdh\"olzli is the central WRRF of the city of Zurich and one of the largest treatment plants in Switzerland, with a design capacity of approximately 670,000 population equivalents. The average flow is on the order of 2,200 L/s under dry-weather conditions, with substantially higher hydraulic loads during storm events. The treatment line comprises mechanical pre-treatment (screening, grit and grease removal, primary sedimentation), followed by a biological activated sludge stage operated as six alternately fed and intermittently aerated lanes (A/I configuration) for combined carbon removal, nitrification and denitrification.

Uster is a medium-sized WRRF with a design capacity of 48,000 population equivalents. It employs conventional mechanical pre-treatment and primary clarification, followed by biological treatment in six parallel sequencing batch reactors (SBRs). The SBRs are operated with alternating phases of filling, aeration, anoxic mixing and settling to achieve carbon removal, nitrification and denitrification.

In this study, these two WRRFs were selected as contrasting full-scale systems to compare microbiome structures under different operating regimes and to relate them to the observed \ce{N2O} emission states.

\subsection{Genomics data acquisition and preprocessing}
Activated sludge biomass was sampled at routine intervals (approximately every 1--2 weeks) from the mixed liquor of the biological reactors at both WRRFs and analyzed by 16S rRNA gene amplicon sequencing to obtain microbial community profiles. For Werdh\"olzli, samples were collected from lane 4 of the biological treatment stage over a period of 1.5 years (December 2021--April 2023), yielding a total of 37 samples. For Uster, previously published 16S rRNA amplicon sequencing data from all six SBR reactors were used, covering the period from September 2018 to May 2019 with a total of 46 samples \citep{gruber2021}.

Raw sequencing reads from Werdh\"olzli were processed using an operational taxonomic unit (OTU)-based workflow. For the Uster dataset, we used the amplicon sequence variants (ASV) as provided in the original publication \citep{gruber2021}, where raw reads had already been processed to infer exact ASVs. For both WRRFs, the resulting OTUs (Werdh\"olzli) and ASVs (Uster) were taxonomically and functionally assigned based on the Microbial Database for Activated Sludge (MiDAS) \citep{dueholm2024}. The exact functional reference that was used in this study can be found in the SI. OTU/ASV counts were then aggregated to the genus level by summing the reads assigned to the same genus, resulting in 2035 detected genera for Werdh\"olzli and 755 for Uster. The time series of genus composition is visualized in the SI. To make samples comparable across time and systems, the resulting genus-level count tables were transformed into relative abundances by dividing each genus count by the total sequencing depth of the corresponding sample. This yielded compositional genus-level community profiles for each sequencing sample, which were used as input for all subsequent microbiome analyses. Since different sequencing protocols were applied to the samples from two WRRFs, we treated the data from them as two separate datasets and analyzed them individually.

\subsection{\ce{N2O} measurement and preprocessing}
\ce{N2O} emissions were monitored at the same reactors and time periods for which microbiome data were available, that is, the lane 4 of the biological activated sludge stage at Werdh\"olzli and all six SBR reactors at Uster. Off-gas measurements were used to quantify \ce{N2O}-N loads at the reactor level, and these data were subsequently processed to derive emission metrics that could be aligned with the microbiome sampling times.

For Werdh\"olzli, the \ce{N2O} emission factor (EF) was calculated by normalizing the off-gas \ce{N2O}-N load to the corresponding influent ammonium load to the biological reactor. Specifically, for each day $t$, the daily emission factor was computed as:
\begin{equation}
    \mathrm{EF}(t) \;=\; \frac{L_{\mathrm{\ce{N2O}\text{-}N}}(t)}{L_{\mathrm{\ce{NH4+}\text{-}N}}(t)\times1.2},
\end{equation}
where $L_{\mathrm{\ce{N2O}\text{-}N}}(t)$ is the daily \ce{N2O}-N emission load and $L_{\mathrm{\ce{NH4+}\text{-}N}}(t)$ is the daily influent \ce{NH4+}-N load (both expressed as kgN/d). The EF value is typically related to the total Kjeldahl nitrogen (TKN) in the influent. Here, we used a coefficient of 1.2 to approximate the TKN influent from the \ce{NH4} loads, as influent TKN measurements were not available. To reduce the influence of short-term operational fluctuations and measurement noise, EFs were aggregated over a 7-day window centered on each microbiome sampling date by averaging the daily EF values within this window. The resulting averaged EF was associated with the corresponding sludge sample. Following the 2019 Refinement to the 2006 IPCC Guidelines \citep{ipcc2006}, an EF threshold of 0.016 was used to categorize each observation into a ``low-emission'' state (EF $<$ 0.016) or a ``high-emission'' state (EF $\geq$ 0.016).

For Uster, detailed influent nitrogen load data at the level of individual SBR reactors were not available. Therefore, only the off-gas \ce{N2O} loads were used as an emission metric, as in \citet{gruber2021}. For each SBR reactor, the 7-day moving average daily \ce{N2O}-N load (kgN/d) was extracted for the days corresponding to microbiome sampling. These loads were then used to assign an emission state to each observation by applying a threshold of 6.2 kgN/d. Samples with a daily \ce{N2O}-N load below this threshold were classified as ``low-emission'', whereas samples at or above 6.2 kgN/d were classified as ``high-emission''. This threshold is the median emission of all samples, ensuring that the numbers of high-emission and low-emission events are equivalent. 

Together, these procedures yielded paired microbiome profiles and binary \ce{N2O} emission states for both WRRFs, forming the basis for the subsequent analysis of archetypal microbiome configurations associated with distinct \ce{N2O} emission regimes.

\subsection{Archetypal analysis of microbiome profiles}
We applied archetypal analysis (AA) to the compositional genus-level abundance data to identify a small set of representative community configurations for each WRRF separately. Each observed community profile is approximated as a convex combination of a limited number of representative community profiles, referred to as ``archetypes''. The archetypes themselves are constrained to be convex mixtures of the observed samples \citep{cutler1994,hart2015}.

Let $x_1,\dots,x_n \in \mathbb{R}^m$ denote the genus-level relative abundance vectors for the $n$ samples from a given WRRF, where $m$ is the number of genera. Because these are compositional data, each $x_i$ has non-negative entries that sum to one. For a fixed number $p$ of archetypes, AA seeks vectors $z_1,\dots,z_p \in \mathbb{R}^m$ such that each archetype is a mixture of the observed samples:
\begin{equation}
    z_k = \sum_{j=1}^n \beta_{kj} x_j, \qquad k=1,\dots,p,
\end{equation}
where
\begin{equation}
    \beta_{kj} \ge 0, \qquad \sum_{j=1}^n \beta_{kj} = 1.
\end{equation}
Each sample $x_i$ should then be reconstructed as a convex combination of the archetypes:
\begin{equation}
    x_i \approx \sum_{k=1}^p \alpha_{ik} z_k,
\end{equation}
where
\begin{equation}
    \alpha_{ik} \ge 0, \qquad \sum_{k=1}^p \alpha_{ik} = 1.
\end{equation}
Combining the two representations gives
\begin{equation}
    x_i \approx \sum_{k=1}^p \alpha_{ik} \sum_{j=1}^n \beta_{kj} x_j.
\end{equation}
The coefficients $\alpha_{ik}$ and $\beta_{kj}$ are estimated by minimizing the residual sum of squares
\begin{equation}
    RSS(p) = \sum_{i=1}^n \left\| x_i - \sum_{k=1}^p \alpha_{ik} z_k \right\|_2^2,
\end{equation}
subject to the non-negativity and unit-sum constraints on $\alpha$ and $\beta$.

We solved the optimization problem using an alternating constrained least squares scheme proposed by \citet{alcacer2024,alcacer2025}. Starting from an initial guess for $\alpha$ and $\beta$, we iteratively updated one matrix while keeping the other fixed. Iterations were continued until the relative change in $RSS(p)$ between consecutive iterations fell below $10^{-4}$ or a maximum of 1000 iterations was reached, whichever occurred first. To avoid local minima, we repeated the optimization with 10 different random initializations of $\alpha$ and $\beta$ for each chosen number of archetypes $p$. For each $p$, the solution with the smallest final $RSS(p)$ was retained as the fitted AA model.

To select an appropriate model complexity, we fitted AA models with $p=1,\dots,7$ archetypes for each WRRF and examined the resulting RSS as a function of $p$. The number of archetypes was chosen based on the smallest $p$ beyond which further increases in $p$ yielded only marginal improvements in the fit. All archetypal analyses were conducted separately for Werdh\"olzli and Uster, using the respective genus-level relative abundance matrices as input. The resulting archetypes and weights were subsequently used to characterize archetypal microbiome configurations and to relate them to the observed \ce{N2O} emission states.

\section{Results and discussion}
\subsection{Three archetypes capture the dominant variation in activated sludge microbiomes}
\begin{figure}[ht!]
    \centering
    \includegraphics[width=0.6\linewidth]{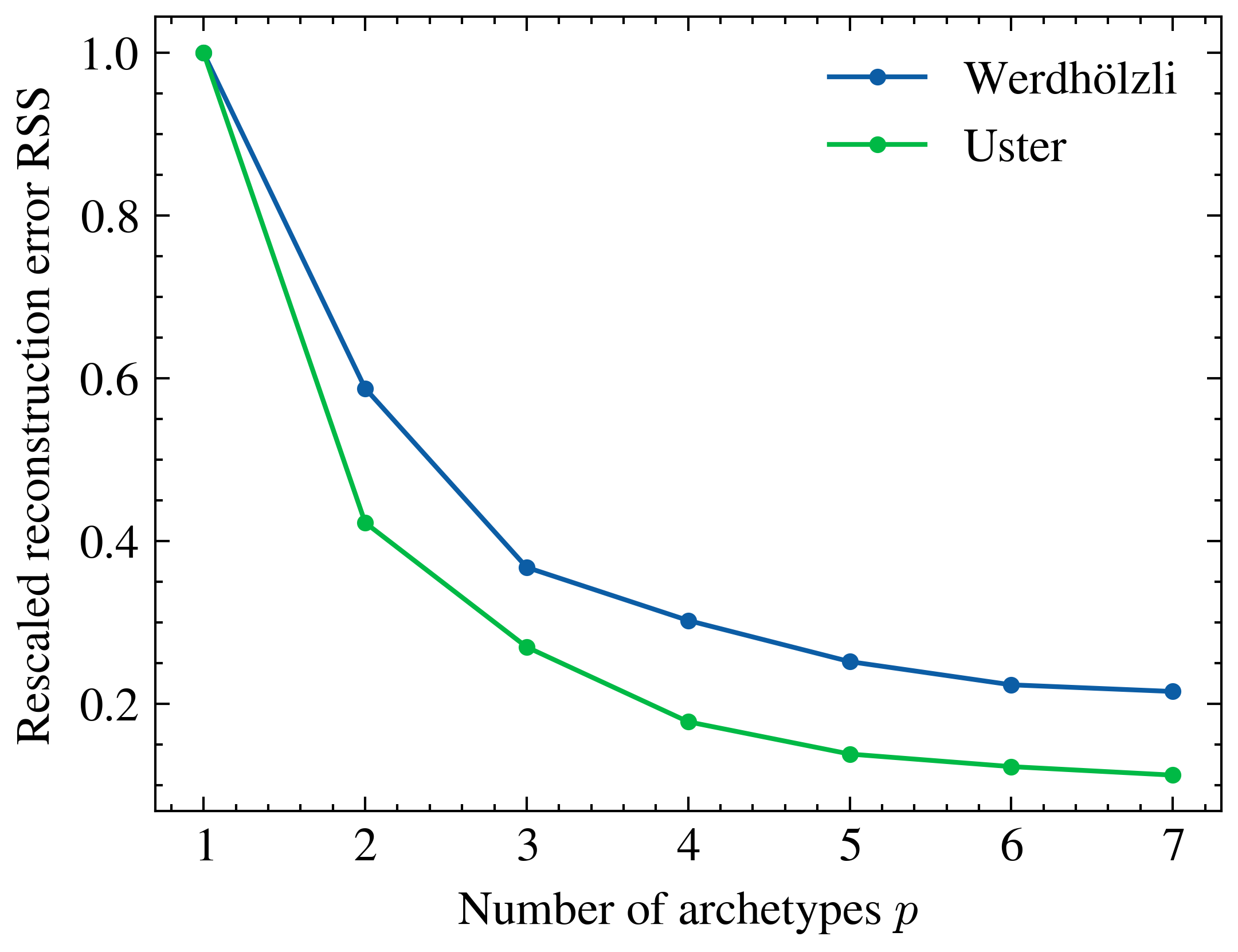}
    \caption{Rescaled reconstruction error (RSS) of the fitted AA model for different numbers of archetypes. The relative RSS for each number of archetypes $p$ is obtained by dividing the absolute RSS by the RSS when $p=1$.}
    \label{fig:RSS_vs_numbers}
\end{figure}

To determine an interpretable yet sufficiently accurate representation of the genus-level community profiles, we fitted AA models with different numbers $p$ of archetypes for each WRRF and evaluated the rescaled reconstruction error (RSS$/\mathrm{RSS}_{p=1}$). As shown in Fig.~\ref{fig:RSS_vs_numbers}, for both Werdh\"olzli and Uster, the reconstruction error decreased steeply when moving from one to a small number of archetypes, followed by a marginal decrease. In particular, increasing $p$ from 1 to 3 yielded the largest gain (Werdh\"olzli: rescaled RSS $\approx 0.37$ at $p=3$; Uster: $\approx 0.27$ at $p=3$). Beyond $p=3$, additional archetypes improved the fit only gradually, indicating that most of the explainable structure in these datasets can be summarized by a small set of extreme community profiles. Therefore, we selected $p=3$ archetypes for both WRRFs for all downstream analyses and named them W1--W3 for Werdh\"olzli and U1--U3 for Uster. This choice provides a compromise between reconstruction quality and interpretability: it captures the majority of the error reduction while keeping the latent state space low-dimensional and therefore is easy to visualize and relate to emission regimes. 

\subsection{Archetypal state space reveals polarized community regimes and aligns with \ce{N2O} emission states}
\begin{figure}[ht!]
  \centering
  \begin{subfigure}{0.48\textwidth}
    \centering
    \includegraphics[height=6cm]{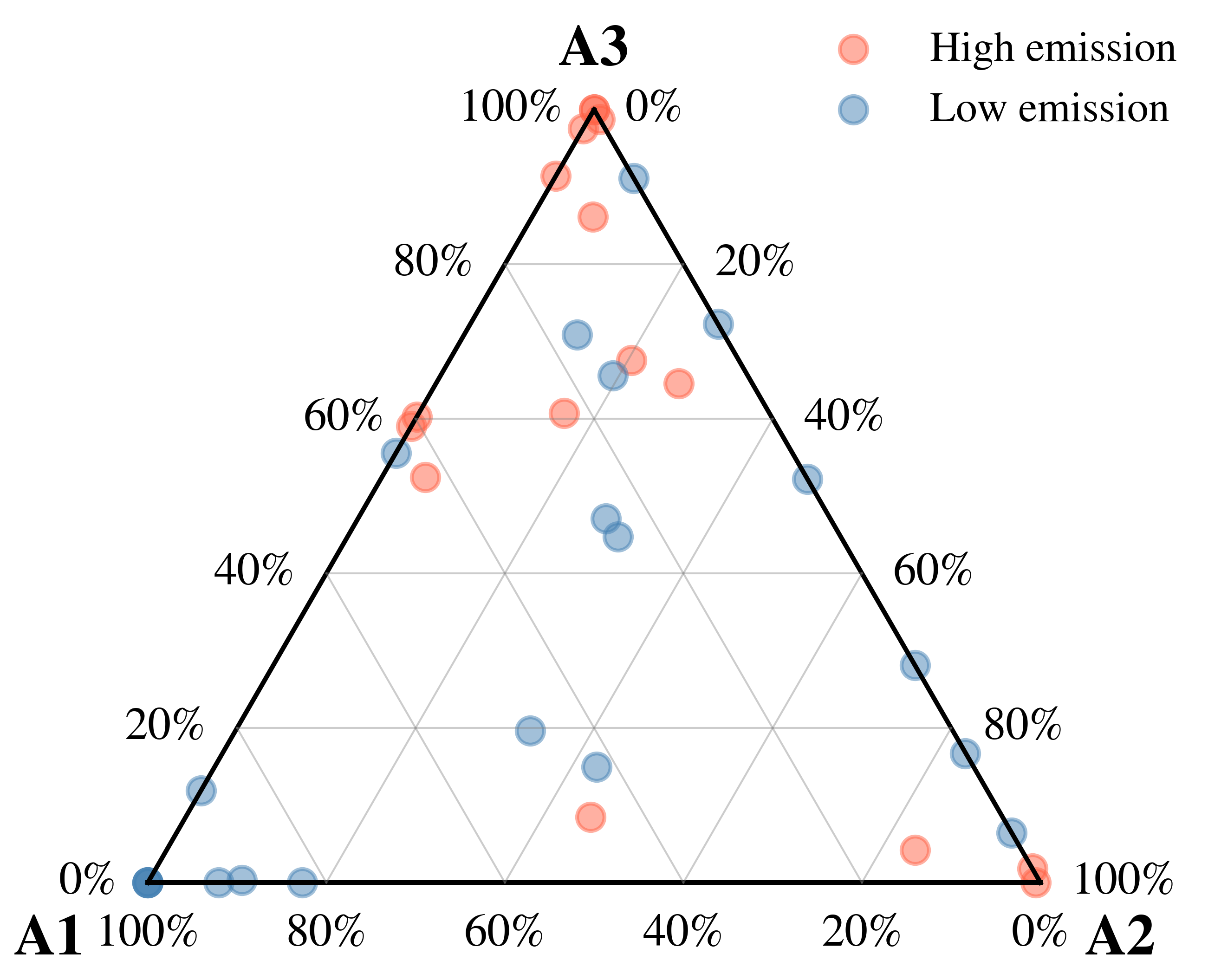}
    \caption{Werdh{\"o}lzli}
    \label{fig:Simplex_W}
  \end{subfigure}
  \hfill
  \begin{subfigure}{0.48\textwidth}
    \centering
    \includegraphics[height=6cm]{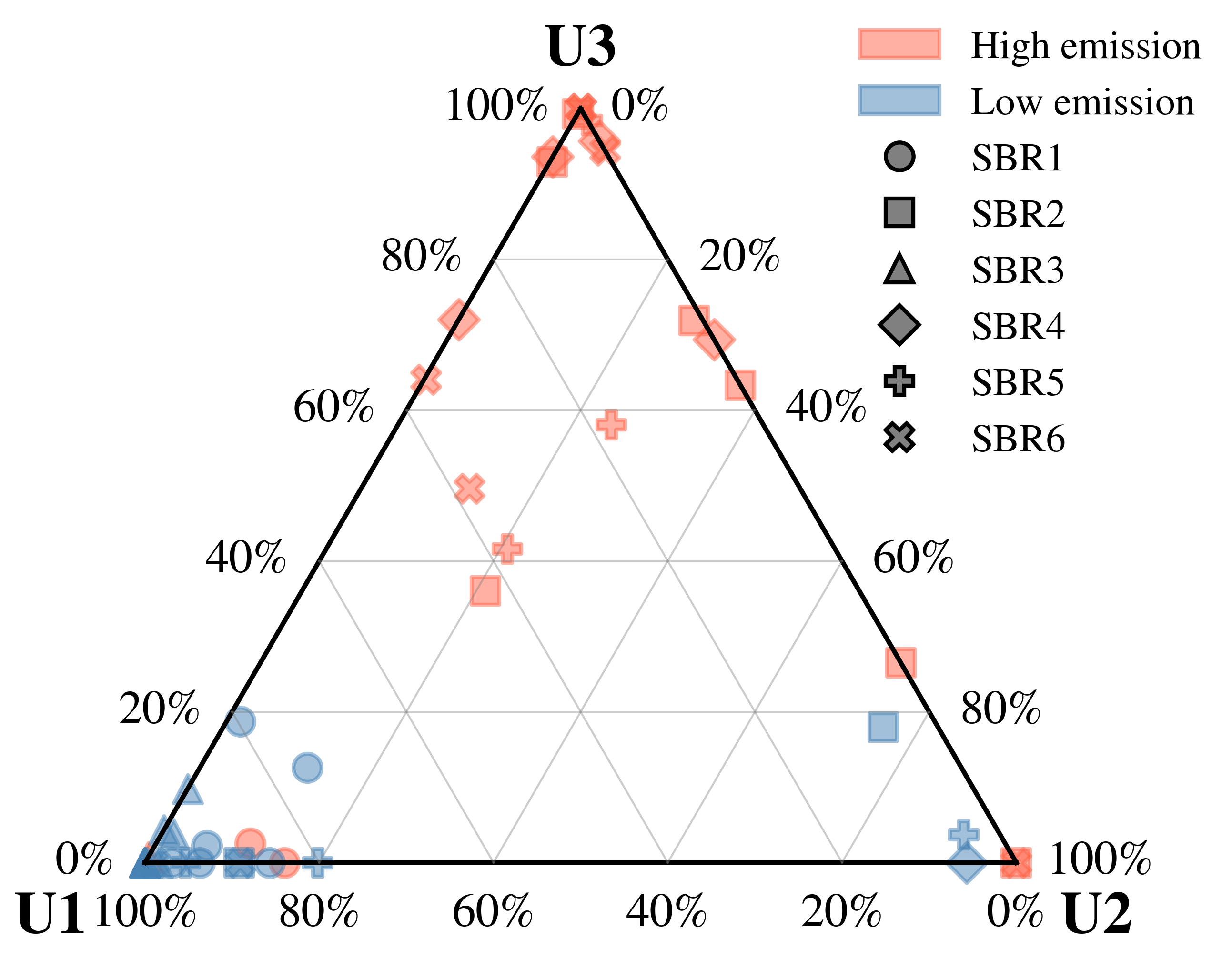}
    \caption{Uster}
    \label{fig:Simplex_U}
  \end{subfigure}
  \caption{Projection of samples on a ternary simplex, the vertices represent pure archetypes. Each sample is constructed by a convex combination of three archetypes. Samples are colored based on the corresponding \ce{N2O} emission state. For Uster, marker shapes denote the individual SBR reactors.}
  \label{fig:Simplex}
\end{figure}

The simplex structure imposed by AA becomes explicit when plotting each sample by its estimated archetype weights in ternary coordinates (Fig.~\ref{fig:Simplex}). By construction, each point represents a convex mixture of three archetypes, so all samples fall inside the triangle whose vertices correspond to pure archetypes. Because the two WRRF datasets differ in sequencing workflow, sampling design and community composition, AA was fitted separately for each plant. Consequently, archetype labels such as W1, W2 and W3 denote plant-specific positions in the respective microbiome state spaces and are not intended to represent one-to-one equivalent community profiles across Werdh\"olzli and Uster. For both WRRFs, many samples lie near the vertices or along the edges, implying that community variability is often dominated by transitions between one or two archetypal profiles rather than requiring substantial contributions from all three simultaneously. This pronounced concentration near the simplex boundaries indicates that the inferred archetypes correspond to functionally relevant and strongly differentiated community regimes.

Although AA was performed without using any emission information, mapping the binary emission labels onto the simplex reveals that microbial composition alone contains a clear signal about \ce{N2O} emission state. At Werdh\"olzli, many low-emission samples cluster predominantly around W1, whereas high-emission samples are enriched toward the W2/W3 region and are especially concentrated near W3 (Fig.~\ref{fig:Simplex_W}). At Uster, the association is even stronger: U1 is primarily occupied by low-emission samples, whereas high-emission samples shift away from U1 and are concentrated mainly toward U3, with some observations extending along the U2--U3 edge (Fig.~\ref{fig:Simplex_U}). The multi-reactor setting at Uster further shows that, despite reactor-specific marker symbols (SBR1--SBR6), high-emission observations from different reactors converge toward the same U3-dominated region of the simplex, suggesting a common ``high-\ce{N2O}'' community configuration across parallel lines. Together, these results show that microbial communities projected in the low-dimensional archetypal state space are strongly polarized, and this polarization aligns with \ce{N2O} emission regimes. The archetype mixture weights can therefore be interpreted as informative microbiome state descriptors that link community composition to reactor performance and can be used to track regime shifts over time.

\subsection{Temporal archetype dynamics track transitions in \ce{N2O} emission states}
\begin{figure}[ht!]
  \centering
  \begin{subfigure}{\textwidth}
    \centering
    \includegraphics[width=\linewidth]{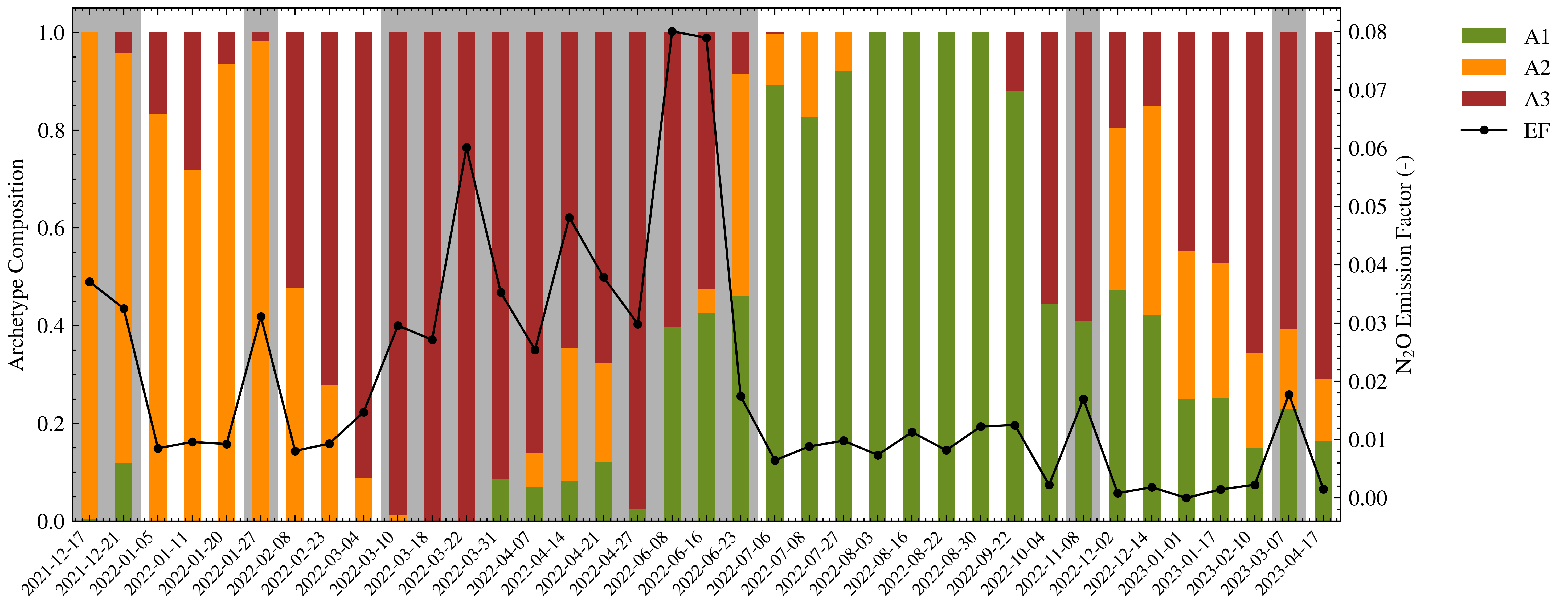}
    \caption{Werdh{\"o}lzli}
    \label{fig:Bar_W}
  \end{subfigure}
  \vspace{1em}
  \begin{subfigure}{\textwidth}
    \centering
    \includegraphics[width=\linewidth]{Figures/bar_U.png}
    \caption{Uster}
    \label{fig:Bar_U}
  \end{subfigure}
  \caption{Archetype compositions (weights) of samples over time. The stacked bars are composed of composition of archetypes. The black lines visualize the corresponding emission states. The gray background indicates the high \ce{N2O} emission periods. The raw archetypal weights are listed in SI.}
  \label{fig:Bar}
\end{figure}

The archetypal representation naturally suits a time-resolved interpretation: each community sampled in a different date can be viewed as a point in a low-dimensional archetypal state space, and microbial community temporal variability corresponds to trajectories in this archetypal space. Fig.~\ref{fig:Bar} visualizes these trajectories as stacked bars of archetype weights aligned with the corresponding \ce{N2O} emission metric. High-emission periods (gray background) therefore highlight time intervals in which the emission threshold was exceeded, enabling a direct comparison between microbiome profile dynamics and shifts in \ce{N2O} emissions.

To summarize this association quantitatively, we computed an archetype-weighted mean emission for each archetype $i$ as
\begin{equation}
    \bar{E}_i = \frac{\sum_t w_i(t)\,E(t)}{\sum_t w_i(t)},
\end{equation}
where $w_i(t)$ is the mixture weight (fraction) of archetype $i$ at time $t$ and $E(t)$ is the corresponding \ce{N2O} emission metric. For both WRRFs, $\bar{E}_3$ was substantially higher than $\bar{E}_1$ and $\bar{E}_2$ (Werdh{\"o}lzli: $\bar{E}_1=0.014$, $\bar{E}_2=0.017$, $\bar{E}_3=0.026$; Uster: $\bar{E}_1=7.08$, $\bar{E}_2=11.52$, $\bar{E}_3=29.70$), providing a compact numerical confirmation that the enrichment in W3 and U3 coincides with elevated \ce{N2O} emissions.

At Werdh\"olzli (Fig.~\ref{fig:Bar_W}), the microbiome exhibited pronounced state switching over the 1.5-year monitoring period. The initial winter phase (December 2021--January 2022) was dominated by W2, with moderate emission factors. In February and March 2022, the community shifted toward a W3-dominated configuration, coinciding with the onset of a sustained high-emission phase. Throughout this spring and early-summer season, W3 remained the dominant contributor for most samples. Notably, this high-emission regime ended with a transition in June--July 2022 to a W1-dominated community; following this shift, the EF dropped below the threshold and remained low for an extended summer period during which W1 was nearly exclusive. In late autumn and winter, the community gradually departed from the stable W1 corner toward mixed compositions with increasing W2 and W3, and sporadic EF spikes into the high-emission category occurred during this drift. Overall, the Werdh\"olzli time series suggests that increases in W3 weight are tightly aligned with the initiation and persistence of high-\ce{N2O} states, whereas sustained W1 dominance corresponds to a low-emission operating regime. Moreover, the succession of W3 from W2 occurs prior to the emission peak, suggesting that the transitions of archetype composition can potentially be trended for early-warning, although a longer validation period is required to corroborate. 

The Uster SBR dataset (Fig.~\ref{fig:Bar_U}) provides an additional test of robustness because it contains six parallel reactors that experience partially synchronized emission dynamics. Across SBR2, SBR4, SBR5, and SBR6, the high-emission phases were consistently characterized by strong enrichment in U3. In contrast, low-emission observations clustered in the U1-dominated state, with U1 occurring prior to the high-emission phase and a recovery trajectory in which U2 replaced U3's share. Notably, SBR1 and SBR3 remained largely U1-dominated across the observation period and exhibited generally low emissions compared to the others, with only limited emission events, indicating the distinct community profiles of these reactors (which is also consistent with their positions near the U1 region of the simplex in Fig.~\ref{fig:Simplex_U}). Despite reactor-to-reactor variability in timing and the presence of intermediate mixtures, the convergence of multiple independent SBR reactors onto a U3-rich configuration during high-\ce{N2O} periods supports the interpretation of U3 as a reproducible microbiome signature of the high-emission state, while U1 represents community regimes associated with low emissions and recovery.

Overall, these temporal patterns strengthen the results from the state space analysis: archetype mixture weights do not only summarize community variation, but also encode dynamic state transitions that coincide with shifts in \ce{N2O} emissions. This temporal view also highlights a practical advantage of AA for process monitoring: approaching microbial profile shifts are visible as a continuous change in archetype weights (particularly the increasing W3/U3 share), suggesting that archetype trajectories could potentially serve as interpretable early-warning indicators for transitions into high-\ce{N2O} emission states.

\subsection{Archetype-specific community signatures suggest \ce{N2O} production and consumption regimes}
\begin{figure}[ht!]
  \centering
  \begin{subfigure}{0.48\textwidth}
    \centering
    \includegraphics[height=10cm]{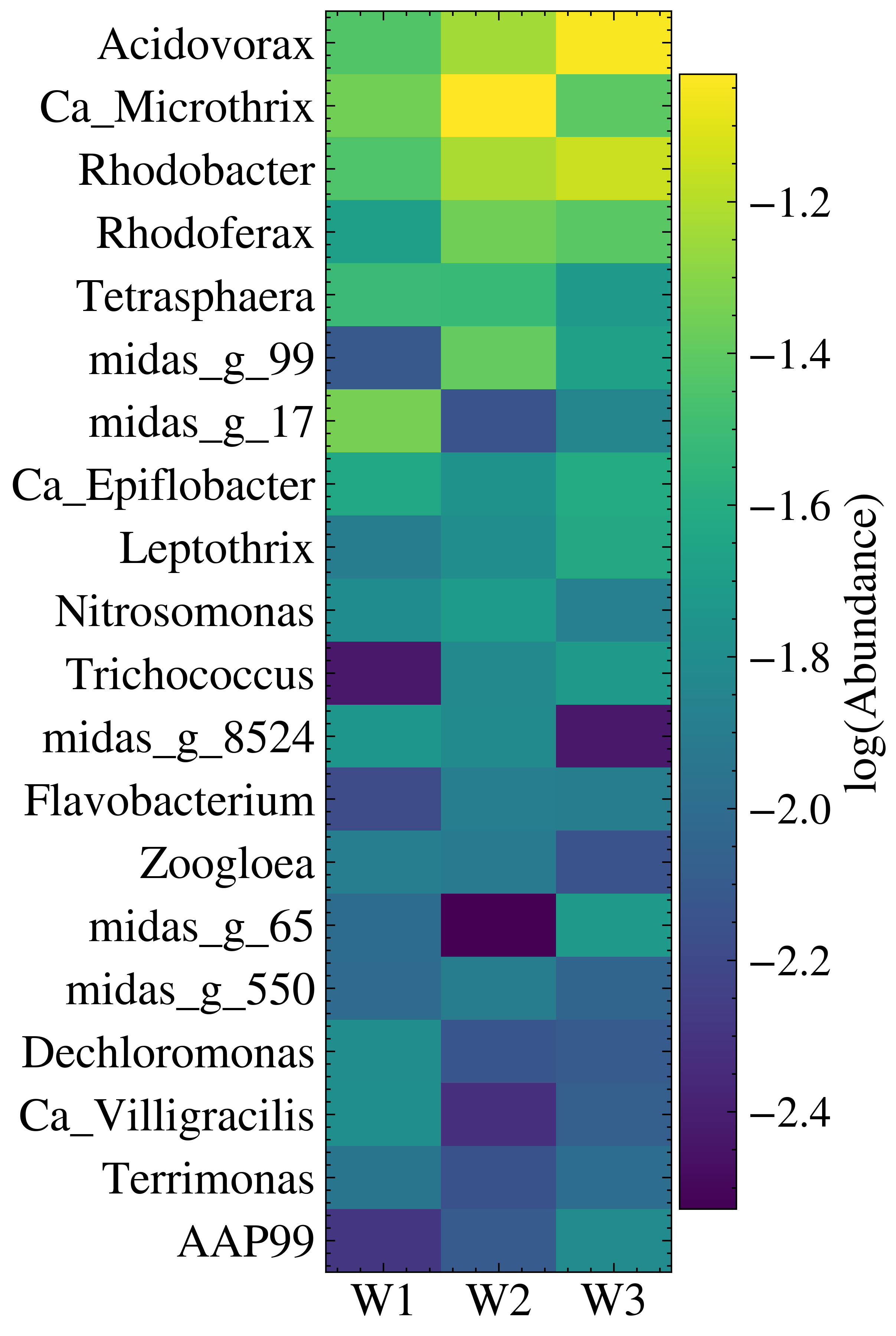}
    \caption{Werdh{\"o}lzli}
    \label{fig:Heat_W}
  \end{subfigure}
  \hfill
  \begin{subfigure}{0.48\textwidth}
    \centering
    \includegraphics[height=10cm]{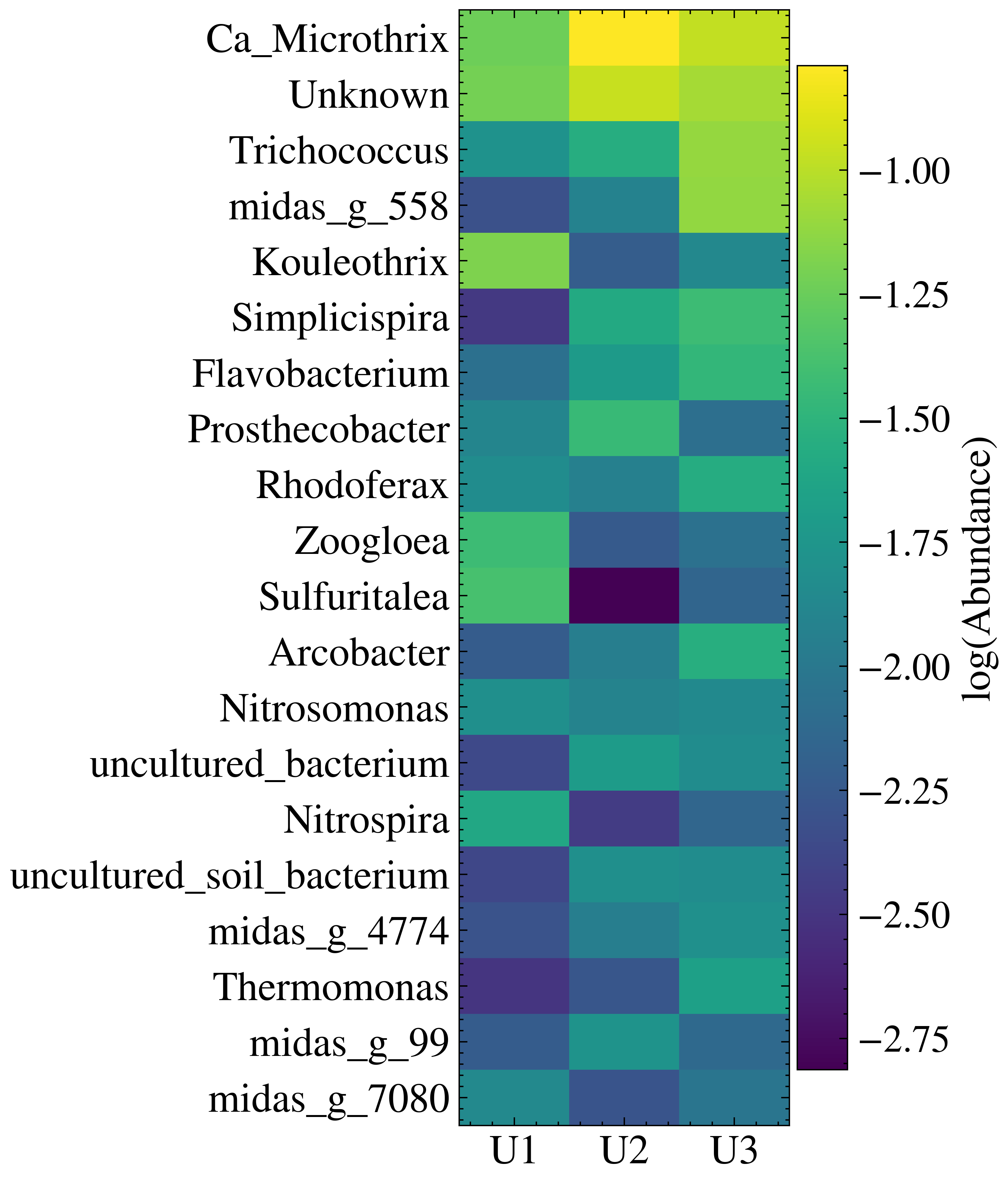}
    \caption{Uster}
    \label{fig:Heat_U}
  \end{subfigure}
  \caption{The composition of genus abundance of archetypes, only the 20 most abundant genera are shown. All abundances are log-transformed. Full genus composition of archetypes can be found in SI.}
  \label{fig:Heat}
\end{figure}

The three archetypes represent markedly different genus-level community configurations rather than subtle perturbations at both WRRFs (Fig.~\ref{fig:Heat}). Because genus-level 16S profiles do not directly quantify pathway activity, we interpret archetype compositions using aggregated functional guilds relevant to \ce{N2O} formation and consumption: AOB, NOB, heterotrophic denitrifiers (DEN), and putative \ce{N2O} reducers (genera reported to carry \textit{nosZ} gene; e.g., \textit{Dechloromonas}, \textit{Flavobacterium}, \textit{Zoogloea} \citep{kim2020,kim2022,qi2022}). Guild sums and derived ratios are summarized in Table~\ref{tab:guilds}. We emphasize that these guilds represent taxonomically inferred potential rather than verified gene presence or expression.

\paragraph{Werdh\"olzli: the high-\ce{N2O} archetype indicates heterotrophic denitrification with reduced \textit{nosZ}-associated signal.}
At Werdh\"olzli, the archetype linked to high emissions (W3) exhibits a pronounced enrichment of denitrifiers compared with the low-\ce{N2O} archetype W1 (DEN: 17.29\% in W3 vs.\ 11.33\% in W1, as shown in Table~\ref{tab:guilds}). In contrast, the summed abundance of \textit{nosZ}-associated denitrifiers decreases from W1 to W3 (3.91\% in W1 to 3.29\% in W3), yielding a strong decrease in the \textit{nosZ}/DEN ratio across archetypes (0.35 in W1, 0.25 in W2, and 0.19 in W3). This indicates that denitrification may not have been fully completed, leading to \ce{N2O} accumulation. Taxonomically, W3 is characterized by high contributions from denitrifier genera (e.g., \textit{Acidovorax} and \textit{Rhodobacter}), while prominent \textit{nosZ}-associated genera such as \textit{Dechloromonas} and \textit{Zoogloea} show lower contributions in W3 than in W1 (Fig.~\ref{fig:Heat_W}). Taken together, this pattern is consistent with an emission regime in which \ce{N2O} production potential via denitrification steps increases without a commensurate increase in \ce{N2O} reduction potential, thereby favoring net \ce{N2O} accumulation. Under this interpretation, heterotrophic denitrification (incomplete reduction from \ce{NO3-}/\ce{NO2-} to \ce{N2}) is hypothesized to be a plausible contributor to the observed high-\ce{N2O} state at Werdh\"olzli, although verification would require gene-resolved evidence (e.g., \textit{nirS/nirK} and clade-resolved \textit{nosZ} quantification) or activity measurements.

\paragraph{Uster: the high-\ce{N2O} archetype aligns with nitrification imbalance, consistent with \ce{NO2-} accumulation.}
At Uster, the most distinctive archetype contrast occurs within the nitrifier guilds. As shown in Table~\ref{tab:guilds}, the low-emission archetype U1 contains a markedly higher NOB (e.g., \textit{Nitrospira}) signal (2.54\%) than U3 (0.71\%) and U2 (0.35\%), while AOB (e.g., \textit{Nitrosomonas}) remains comparatively similar across archetypes (1.43--1.77\%). As a result, the AOB/NOB ratio increases from 0.70 (U1) to 2.20 (U3), with U2 representing an extreme state (4.09) that is rarely realized as a pure community (see Fig.~\ref{fig:Simplex_U}). This pattern indicates a shift toward reduced nitrite-oxidation capacity relative to ammonia-oxidation in the high-\ce{N2O} archetype, a community-level signature consistent with \ce{NO2-} accumulation and nitrification failure scenarios linked to seasonal \ce{N2O} peaks in the same SBR system \citep{gruber2021}. However, the emission-associated archetype cannot be identified from the AOB/NOB ratio alone. Although U2 has the highest AOB/NOB ratio, high-emission samples are not concentrated exclusively at the U2 vertex; instead, they are most consistently enriched toward U3, with some observations distributed along the U2--U3 edge (Fig.~\ref{fig:Simplex_U}). Thus, U2 likely represents a rarely extreme nitrifier-imbalanced or transitional state, whereas U3 represents the more reproducible high-emission community regime. This interpretation is also consistent with the temporal dynamics (Fig.~\ref{fig:Bar_U}), where high-\ce{N2O} periods across several SBRs coincide most clearly with increased U3 weights. Therefore, U3 should be interpreted as the high-\ce{N2O} archetype not because it has the maximum AOB/NOB ratio, but because it combines a nitrification-imbalance signal with the strongest empirical association with elevated \ce{N2O} emissions in both the simplex and time-series analyses.

Across the two WRRFs, the high-\ce{N2O} archetypes (W3/U3) were not defined by the same genera or by the same inferred pathway signature. This is expected because the AA models were fitted separately for each plant, and the two datasets differed in sampling design, sequencing workflow, process configuration and microbial community composition. Therefore, archetypes should be interpreted as plant-specific signatures rather than directly equivalent community states across WRRFs. Within each plant, however, the archetype associated with high emissions could be linked to a pathway-relevant functional interpretation. At Werdh\"olzli, W3 was characterized by increased denitrifier abundance together with a lower \textit{nosZ}-associated fraction, suggesting that incomplete denitrification or reduced \ce{N2O}-sink capacity may contribute to net \ce{N2O} accumulation. At Uster, U3 was instead characterized by reduced NOB abundance relative to AOB, consistent with the previously reported \ce{NO2-}-accumulating, high-\ce{N2O} state in this SBR system \citep{gruber2021}. Thus, the common result across plants is not a shared taxonomic or pathway signature, but the fact that AA identified a plant-specific community regime associated with elevated \ce{N2O} emissions. A joint cross-plant AA could in principle be used to test whether shared archetypes exist across WRRFs, but this was not the objective of the present analysis and would require harmonized preprocessing and careful treatment of plant-specific sequencing and compositional effects. Here, the separate analyses were chosen to determine whether interpretable low-dimensional microbiome states can be identified within each WRRF and related to its own emission dynamics.

\begin{table}[ht!]
\centering
\caption{Functional guild summaries of archetype compositions. Values are summed relative abundances (\%) of genera assigned to each guild; AOB/NOB ratio and \textit{nosZ}-associated guild/denitrifiers ratio are calculated.}
\label{tab:guilds}
\begin{tabular}{cccccccc}
\toprule
\multicolumn{2}{l}{} & \multicolumn{1}{c}{AOB (\%)} & \multicolumn{1}{c}{NOB (\%)} & \multicolumn{1}{c}{AOB/NOB} & \multicolumn{1}{c}{Denitrifiers (\%)} & \multicolumn{1}{c}{\textit{nosZ} (\%)} & \multicolumn{1}{c}{\textit{nosZ}/DEN} \\
\midrule
\multicolumn{8}{l}{\textbf{Werdh\"olzli}} \\
\midrule
 & W1 & 2.18 & 1.36 & 1.60 & 11.33 & 3.91 & 0.35 \\
 & W2 & 2.50 & 1.92 & 1.30 & 14.77 & 3.64 & 0.25 \\
 & W3 & 1.84 & 2.01 & 0.92 & 17.29 & 3.29 & 0.19 \\
\midrule
\multicolumn{8}{l}{\textbf{Uster}} \\
\midrule
 & U1 & 1.77 & 2.54 & 0.70 & 13.36 & 5.67 & 0.42 \\
 & U2 & 1.43 & 0.35 & 4.09 & 8.24 & 4.05 & 0.49 \\
 & U3 & 1.57 & 0.71 & 2.20 & 12.58 & 5.62 & 0.45 \\
\bottomrule
\end{tabular}
\end{table}

\subsection{Temperature regulates the archetypal state space and likely mediates seasonal \ce{N2O} emission}
\begin{figure}[ht!]
  \centering
  \begin{subfigure}{\textwidth}
    \centering
    \includegraphics[width=\linewidth]{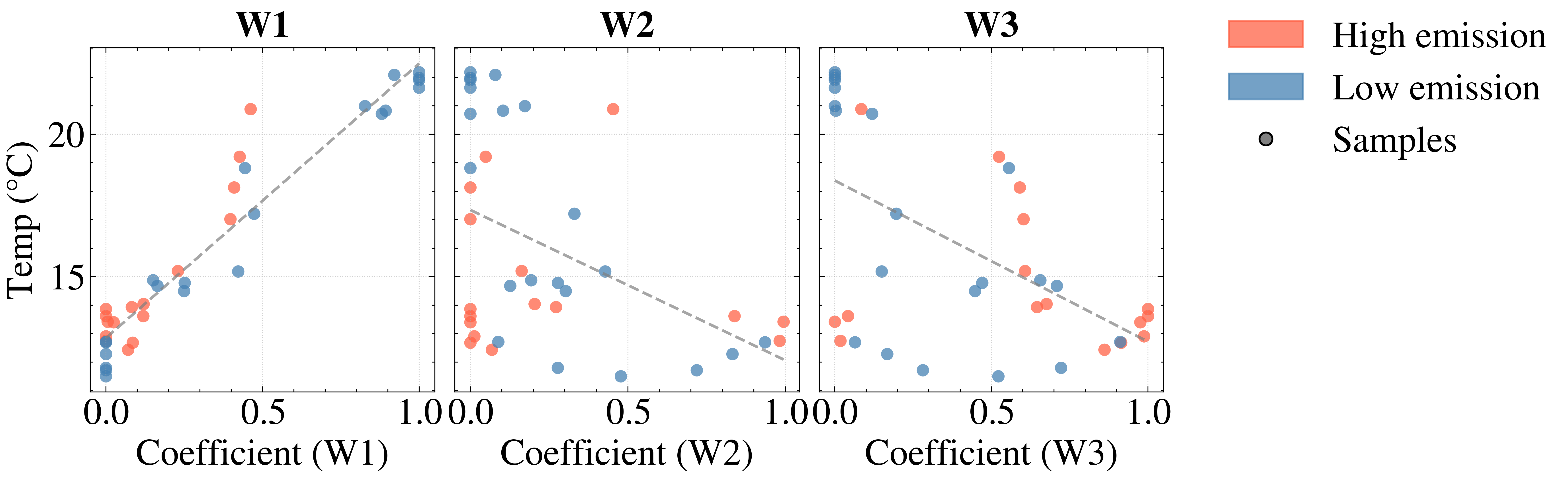}
    \caption{Werdh{\"o}lzli}
    \label{fig:Temp_weight_W}
  \end{subfigure}
  \vspace{1em}
  \begin{subfigure}{\textwidth}
    \centering
    \includegraphics[width=\linewidth]{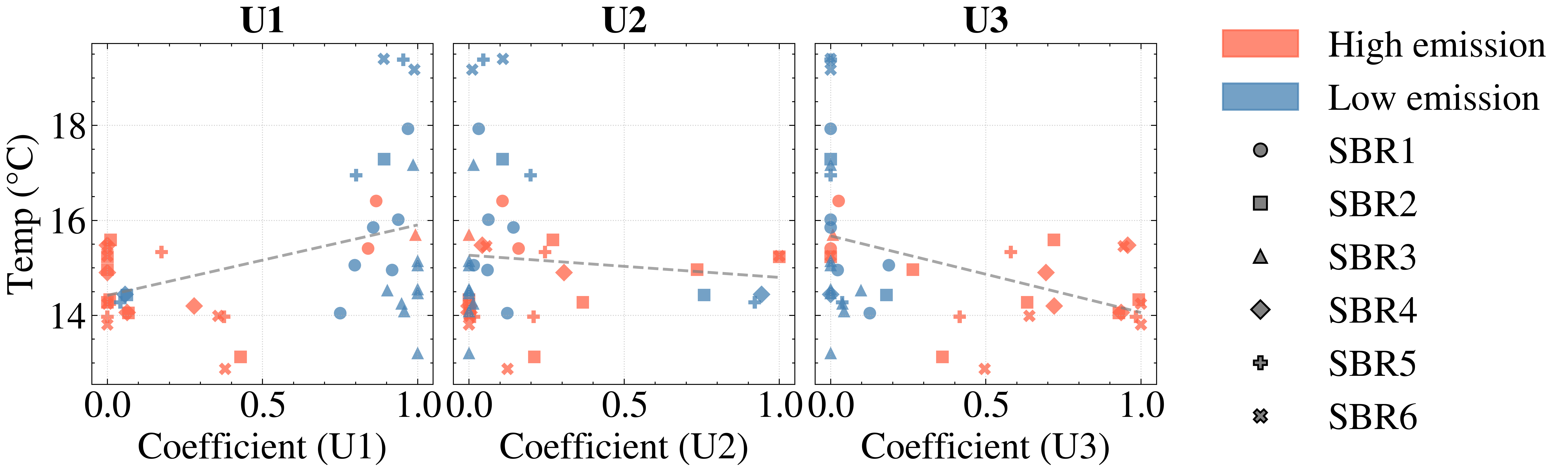}
    \caption{Uster}
    \label{fig:Temp_weight_U}
  \end{subfigure}
  \caption{Relationship between influent wastewater temperature and archetype coefficients. Each point represents one sample. Colors indicate \ce{N2O} emission state. For Uster, marker shapes denote the individual SBR reactors. Dashed lines show linear trends between temperature and archetype weights.}
  \label{fig:Temp_weight}
\end{figure}

To evaluate how temperature relates to the archetypal microbiome state space, we examined the correlation between influent wastewater temperature and the sample-specific archetype coefficients, while simultaneously indicating the observed \ce{N2O} emission state (Fig.~\ref{fig:Temp_weight}). Because AA represents each sample as a convex mixture of three archetypes, these plots directly show how seasonal forcing shifts the community toward or away from specific archetypal regimes. In both WRRFs, temperature is strongly associated with the archetype coefficients, indicating that a substantial part of the microbiome variation captured by AA follows a seasonal axis, consistent with previous reports on temperature-dependent \ce{N2O} dynamics and microbial community shifts in activated sludge \citep{guo2010,poh2015,daelman2015,bao2018,gruber2020,wang2020,gruber2021,valk2022,roothans2025}.

At Werdh\"olzli (Fig.~\ref{fig:Temp_weight_W}), the W1 coefficient increases strongly with temperature, whereas the weights of W2 and W3 decrease with temperature. Thus, warm conditions are associated with W1-dominated communities, while colder conditions shift the microbiome toward W2- and especially W3-enriched states. This pattern matches the emission stratification observed earlier: low-emission samples are concentrated at high W1 and low W2/W3 weights, while many high-emission samples occur at intermediate-to-high W3 weights and lower temperatures. The relationship is not perfectly deterministic, but the overall correlation structure indicates that temperature acts as an upstream driver that reorganizes the microbiome within the archetypal state space, thereby altering the likelihood of entering the high-\ce{N2O} regime.

At Uster (Fig.~\ref{fig:Temp_weight_U}), a similar overall seasonal organization is observed, although the multi-reactor setting adds some scattering. U1 shows a clear positive association with temperature, while U2 and U3 are negatively associated with temperature. High-temperature observations are therefore concentrated in U1-dominated samples, whereas colder conditions favor larger U2 and U3 weights. Importantly, high-emission observations across several SBRs cluster predominantly at high U3 weights and low temperatures, while low-emission samples are more common at high U1 weights, supporting the idea of two common contrasting microbiome configurations: a cold, high-\ce{N2O} one and a warm, low-\ce{N2O} one.

Across both WRRFs, these relationships between archetype coefficients and temperature explicitly show which archetypal regimes are favored under warm versus cold conditions and how these shifts intersect with emission states. Because the high-\ce{N2O} archetype region is primarily occupied during colder conditions, a simple hypothesis is that temperature not only covaries with emissions, but also mediates seasonal changes in microbiome composition that increase the likelihood of functional imbalances that favor net \ce{N2O} accumulation (e.g., inhibition or loss of NOB relative to AOB \citep{yao2017,gruber2021}, or incomplete denitrification relative to \ce{N2O} production \citep{law2012}). Although temperature is clearly associated with both archetypal microbiome composition and \ce{N2O} emission state, visual inspection and exploratory comparisons indicated that temperature alone did not separate high- and low-emission observations as clearly as the archetype coefficients. We therefore interpret temperature as an upstream seasonal driver that shifts the microbiome toward or away from emission-associated archetypal states, rather than as a substitute for the microbiome-based state representation. From an operational perspective, monitoring archetype weights together with temperature may therefore provide a mechanistically interpretable indicator of seasonal transitions toward high-\ce{N2O} emission states.

\section{Conclusions}
This study shows that archetypal analysis (AA) provides an interpretable, compositional representation of activated sludge microbiomes that aligns strongly with \ce{N2O} emission regimes at full-scale WRRFs. Using genus-level 16S profiles paired with \ce{N2O} emission metrics from two different WRRFs, three archetypes are sufficient to capture the dominant community variability and define a low-dimensional simplex state space in which samples are concentrated near vertices and edges, indicating polarized community configurations. Although the AA model is trained without emission information, a consistent high-emission region emerges in both systems: high-\ce{N2O} observations are characterized by increased weights of a specific archetype (W3/U3), and temporal trajectories show that shifts toward this archetype coincide with the onset and persistence of high-emission periods.

Archetype-specific functional guild summaries suggest that similar emission states can arise from different site-specific community configurations. At Werdh\"olzli, the high-\ce{N2O} archetype is associated with an increased denitrifier signal, together with a reduced \textit{nosZ}-associated denitrifier fraction, consistent with incomplete denitrification and reduced \ce{N2O}-sink capacity. At Uster, the high-\ce{N2O} archetype exhibits a reduced NOB abundance relative to AOB, consistent with nitrification imbalance and \ce{NO2-} accumulation. Temperature structures the archetypal state space at both sites, indicating that seasonal forcing likely mediates shifts toward microbiome configurations with higher \ce{N2O} accumulation potential.

Overall, AA yields operationally meaningful microbiome state variables that can be monitored over time and may serve as explainable early-warning indicators for transitions into high-\ce{N2O} emission regimes. Future works should validate these compositional indicators using gene-resolved or activity-based measurements (e.g., quantification of functional genes \textit{nosZ}, \textit{nirK} and \textit{nirS}; metatranscriptomics/proteomics; and \textit{ex situ} kinetics), and evaluate whether integrating archetype trajectories with routinely measured operational variables can improve predictive control strategies for \ce{N2O} mitigation across diverse WRRF configurations. It is also worth noting that the aim of this work is to investigate the qualitative relationship between low-dimensional microbiome representations and \ce{N2O} emissions. The quantitative \ce{N2O} prediction based on variate inputs, including operational parameters, raw abundance profiles and archetypal weights, is not within the scope of this work.

\section*{Code and data availability statement}
The scripts and data used in this study are available at: \url{https://github.com/cchen07/Archetype_N2O}

\section*{CRediT authorship contribution statement}

Cheng Chen: Conceptualization, Data curation, Formal analysis, Methodology, Software, Visualization, Writing -- original draft.

Marcello Seppi: Methodology, Writing -- review \& editing.

Samir Suweis: Methodology, Writing -- review \& editing.

Andreas Froemelt: Funding acquisition, Methodology, Project administration, Writing -- review \& editing.

Eberhard Morgenroth: Methodology, Supervision, Writing -- review \& editing.

Andreas Scheidegger: Methodology, Writing -- review \& editing.

Carlo Albert: Conceptualization, Funding acquisition, Methodology, Project administration, Supervision, Writing -- review \& editing.

\section*{Declaration of competing interest}
The authors declare that they have no known competing financial interests or personal relationships that could have appeared to influence the work reported in this paper.

\section*{Acknowledgments}
This research was supported by the Swiss National Science Foundation (SNSF) under Grant No. 219321. The authors gratefully acknowledge the financial support provided for this work.

\section*{Declaration of generative AI use}
During the preparation of this work, the authors used AI to assist with language polishing and result visualization. After using this tool, the authors reviewed and edited the content as needed and took full responsibility for the content of the publication.

\bibliographystyle{apalike}
\bibliography{references}

\end{document}